# Lithium borohydride ($LiBH_4$): An innovative material for neutron radiation shielding

Mohammadreza Lotfalian, Mitra Athari Allaf*, Masoud Mansouri
*Department of Nuclear Engineering, Science and Research Branch, Islamic Azad University, Tehran, Iran*

HIGHLIGHTS

- The inclusion of lithium borohydride resulted in an enhancement of shield efficiency.
- The performance of lithium borohydride surpassed that of boron carbide in terms of effectiveness against fast neutrons.
- Recent calculations of the shielding data were conducted and compared with earlier shield configurations.
- The longevity of the engineered shield is deemed adequate for its intended application.

ABSTRACT

Radiation shielding plays a crucial role in various industries, including nuclear and space exploration. Among the most abundant elements and isotopes found in nature, B-10 has one of the highest neutron absorption cross-sections, closely followed by Li-6. It is worth noting that hydrogen, with its light nucleus, serves as an excellent neutron reflector. Surprisingly, the potential of the lithium borohydride molecule ($LiBH_4$), which consists exclusively of these elements, as a shield against neutron radiation has not yet been explored. This study investigates various materials that can potentially be used as shields. First, we assessed traditional shields and previous optimizations for shielding. The findings showed that concrete containing 10% $B_4C$ yielded the best results. High-performance concrete (HPC) replaced regular concrete. By gradually incorporating lithium borohydride into the shield, along with the appropriate level of boron carbide, further optimization was achieved. Calculations were performed using the MCNPX 2.7E code. The introduction of the new shield resulted in a significant 40% reduction in volume compared with the previous sample. The study findings showed that a 30 cm thick shield effectively blocked 95% of the total neutrons and 92% of the total gamma radiation. Additionally, it was noted that the shielding effects of lithium borohydride against fast neutrons are greater than those of boron carbide. Various parameters and data of the designed shield were calculated and compared with those of the previous sample.

## 1 Introduction

Safety and health have always been primary concerns, leading researchers to continuously develop nuclear shielding materials for several decades. However, limitations in volume and transportation of these shields have prompted researchers to explore alternative materials. It is predicted that in the future, there will be a greater focus on newer and more accessible materials that occupy less space. One such material is high performance concrete (HPC), which can be produced using conventional materials and processes. In addition, neutron shielding requires the use of boron and lithium hydride (LiH) as filler materials. Specifically, natural boron consists of 19.8% of B-10 (Takasaki et al., 2012) and natural lithium consists of 7.59% of Li-6 (Scott and Shafer, 2020). Considerable research has already been conducted on shield optimization. Some researchers have focused on optimizing the proportion of metals in the alloy, whereas others have explored the use of fillers in concrete or polyethylene.

A research investigation focused on compounds characterized by their significant neutron absorption capabilities, attributed to their elevated absorption cross-section. These compounds were incorporated in varying volume proportions within a cubic framework, utilizing concrete as the primary substrate for radiation shielding. To achieve the desired volume percentages, the diameters of the absorbent spheres were adjusted, ranging from 200 $\mu$m (1%) to 750 m (52%). The constructed sample was subjected to exposure from an Am-Be neutron source with an intensity of 5 Ci, of which 37.5% was attributed to gamma radiation. The study examined the compounds cadmium oxide (CdO), boron carbide ($B_4C$), and boron nitride (BN), which were integrated into serpentine concrete. Simulation outcomes indicated that, considering the efficiency of the various compounds, the feasibility of their production, and the associated costs, both boron carbide and boron nitride demonstrated commendable effectiveness in minimizing equivalent doses (Alipour et al., 2019).

*Corresponding author: athari@srbiau.ac.ir

A group of researchers focused on enhancing the weight and dimensions of traditional $^{241}$Am-Be shield designs by employing cutting-edge advanced materials. The incorporation of increased hydrogen and/or carbon content in these neutron-absorbing materials significantly improves their ability to slow down neutrons when compared to conventional alternatives. The first design utilizes a combination of "Magnesium Borohydride" and "Kennertium" for effective neutron and gamma absorption, resulting in a 60% reduction in volume and a 27% reduction in weight. In the second design, "Kennertium" was substituted with a "Polyethylene-78.5% Bismuth" alloy, leading to a significantly lighter shield configuration, with reductions in volume and weight of 49% and 67%, respectively (Moadab and Saadi, 2019).

A recent investigation has introduced a novel nano polymer particle that incorporates both gamma and neutron absorbers, aimed at enhancing the simplicity and compactness of shield design. This synthesized polymer utilizes $H_3BO_3$ as the neutron absorber and PbO as the gamma ray attenuator, resulting in the formation of PbO-$H_3BO_3$ nano polymer particles. The findings indicated that the incorporation of 5 wt% PbO-$H_3BO_3$ led to a 38% reduction in the half-value layer (HVL) at 511 keV and a 24% reduction at 1332 keV. Furthermore, the results demonstrated that the use of nano-sized particles offers significant advantages over their micro-sized counterparts in terms of neutron shielding efficacy (Mokhtari et al., 2021b).

A study has focused on investigating the shielding characteristics of conventional concrete that is reinforced with varying weight fractions of lead oxide micro/nano particles. The findings indicated that the incorporation of lead oxide filler particles significantly improves the mass attenuation coefficient of the conventional concrete. Specifically, an increase in the weight fraction of nanoparticles corresponded to a rise in the mass attenuation coefficient. Experimental data revealed that the half-value layer (HVL) of the conventional concrete reinforced with 5 wt% of nano PbO particles was diminished by 64% at an energy level of 511 keV and by 48% at 1332 keV. A satisfactory correlation was observed between the simulation and experimental outcomes, demonstrating that the addition of nano PbO particles is particularly effective at lower gamma energies, up to 1 MeV. Conversely, there were negligible differences noted between the effects of micro and nano-sized particles (Mokhtari et al., 2021a).

A study was conducted to develop HPC samples that are both mechanically strong and durable. This study focused on analyzing the mechanical properties, crack resistance, sulfate attack resistance, frost resistance, and impermeability of concrete with various mineral admixtures, specifically mineral powder and fly ash. Moreover, considering the aspect of sustainability and future trends in the advancement of HPC will offer a more holistic outlook within this domain (Han and Zhou, 2023). Findings from an investigation showed that high concentrations of tungsten oxide ($WO_3$) in newly developed polymers can decrease their mean free path values. The results also indicated that the number of photons that penetrated depended on their energy (Elsafi et al., 2023).

In a study a composite shielding material was created using a hot-pressure sintering method. The material consisted of 10.00 wt% boron carbide particles ($B_4C$), 13.64 wt% surface-modified cross-linked polyethylene (PE), and 76.36 wt% tungsten particles. The results demonstrated the composite's effectiveness in shielding high-energy neutrons. Simulation tests conducted in a white neutron field showed a remarkable 99.99% reduction in fast neutron penetration after a shielding thickness of 44 cm. Additionally, experimental results indicated a 99.70% decrease in neutron penetration (Shao et al., 2024).

In an innovative research project, researchers tested and studied the effect of PbO-$H_3BO_3$ polymer nanoparticles in ordinary concrete. The results showed that nanoparticles performed better than microparticles in optimizing the shield against neutron radiation. Additionally, both modes improved the performance of the shield equally in terms of shielding against gamma radiation (Mokhtari et al., 2021b). In a study on particle behavior, four commonly used algorithms (SCE, TLBO, DE, and GA) were used to optimize shield materials. The researchers concluded that the SCE algorithm performed the best. They then used this algorithm to optimize the shield compound and ultimately selected a multilayer composite (Cai et al., 2018). In another study, the impact of tungsten oxide microparticles and nanoparticles on the concrete attenuation coefficient was investigated using the MCNPX2.4.0 code. The study found that adding $W_3O$ microparticles, particularly $W_3O$ nanoparticles, strengthened the shield against gamma rays (Tekin et al., 2017). In a separate study, the polyethylene compound was optimized for shielding against gamma and neutron beams by adding boron carbide and tungsten. The study concluded that adding 5 wt% of boron carbide to polyethylene effectively absorbed thermal neutrons, whereas adding 45 wt% of tungsten to polyethylene effectively absorbed gamma radiation (Salimi et al., 2018).

In one study, an alloy made from different ratios of aluminum, iron, copper, and lead was tested against gamma rays. The compound containing 5% Al, 40% Fe, 50% Cu, and 5% Pb was selected as the optimal ratio after simulating four different ratios (Reda, 2016). In another study, researchers analyzed different compounds of concrete and boron carbide as shields against gamma radiation. Numerical simulations demonstrated that adding boron carbide to concrete effectively enhanced its ability to shield against gamma radiation (Stanković et al., 2010). In a new and intriguing study, researchers analyzed specific hexaborides used as gamma radiation shields. The researchers experimentally synthesized the samples and then determined the chemical and physical characteristics of the manufactured samples. Ultimately, they concluded that the manufactured hexaborides show promising results compared with other gamma radiation shields that have been investigated in previous studies (Gulbicim et al., 2023). In a study, researchers examined concrete containing magnetite and limonite with different ratios as a shield against gamma rays. The study showed that adding limonite and, especially, magnetite, effectively improved the gamma attenuation coefficient of concrete (Gur et al., 2017).

In an investigation, a shielding system for Nuclear-pumped lasers (NPL) designed that uses $^{10}B(n,\alpha)^{7}Li$. The researchers used combinations with different materials with various arrangements in three layers. According to the simulation, the arrangement of $Fe_2B$-BPE-Pb is a suitable protection compound for such lasers. They used MCNPX 2.6.0 Monte Carlo code and the thermal neutron flux as $1 \times 10^{16}$ $n.cm^{-2}.s^{-1}$ for excitation reaction (Shamsinasab et al., 2024).

One study aimed to evaluate the shielding performance of ethylene propylene diene monomer (EPDM) rubber composites filled with 200 PHR (parts per hundred of rubber) of different metal oxides (either $Al_2O_3$, CuO, CdO, $Gd_2O_3$, or $Bi_2O_3$) as protective materials against gamma and neutron radiations. For this purpose, different shielding parameters, including the linear attenuation coefficient ($\mu$), mass attenuation coefficient ($\mu/\rho$), mean free path (MFP), half value layer (HVL), and tenth value layer (TVL), were calculated. Based on $\mu/\rho$ values, other significant shielding parameters such as the effective atomic number ($Z_{eff}$), effective electron density ($N_{eff}$), equivalent atomic number ($Z_{eq}$), and exposure buildup factor (EBF) were also computed. This study demonstrates that the gamma-radiation shielding performance of the proposed metal oxide/EPDM rubber composites is increasing in the order of EPDM < $Al_2O_3$/ EPDM < CuO/ EPDM < CdO/EPDM < Gd2O3/ EPDM < $Bi_2O_3$/ EPDM. Furthermore, three sudden increases in the shielding capability in some composites occur at 0.0267 MeV for CdO/EPDM, 0.0502 MeV for $Gd_2O_3$/ EPDM, and 0.0905 MeV for $Bi_2O_3$/ EPDM composites (Alabsy and Elzaher, 2023).

The findings of the Radiation Assessment Detector (RAD) revealed that a journey to Mars would expose astronauts to a significant amount of harmful radiation. While previous robotic missions to Mars have not been successful, future human explorers heading to the Red Planet should be aware of the potential risks associated with radiation in deep space. The measurements obtained by the Curiosity rover during its voyage to Mars in August 2022 indicate that the impact of radiation on human space travelers remains a significant concern, although it is not yet fully understood (Kerr, 2013). A recent study introduced a super lattice nano-barrierenhanced carbon fiber reinforced polymer (CFRP) with a density of approximately 3.18 $g.cm^{-3}$. This innovative material seamlessly integrates with the mechanical properties of CFRP, effectively becoming a part of the composite structure. The research findings emphasize the necessity of both external and internal shielding mechanisms to safeguard satellites from trapped protons and electrons. Charged particles, such as protons with energies ranging from 0.1 to 400 MeV, are ensnared by the planet's robust magnetic field, forming radiation belts. These belts consist of an inner zone located between 6000 and 12,000 km altitude and an outer zone spanning from 25,000 to 45,000 km altitude, directly impacting satellites in orbit (Delkowski et al., 2023).

In one investigation, the use of celestite ($SrSO_4$) minerals as aggregates in barrier composites was explored to ensure reliable handling in repositories, radiotherapy rooms, and research centers constructed with cement-based composites. The favorable shielding properties of celestite minerals make them ideal for this purpose. This study thoroughly examined the high-rate X-ray shielding ability and mechanical performance of the developed real radiation scenarios. Additionally, the researchers found that the interface properties between the composite paste and celestite minerals remained compatible even when up to 30% of the celestite aggregate was replaced (Öztürk et al., 2022).

In the quest for the advancement of environmentally sustainable nuclear technology, scientists are discovering innovative ways to repurpose nuclear waste to create biodiesel. Therefore, it is imperative to enhance radiation protection protocols in tandem with the evolution of nuclear technology (Devarajan, 2024).

A research study was conducted to experimentally investigate the shielding effectiveness of nitrile butadiene rubber (NBR) against neutrons and gamma rays using the beam line of the Isfahan Miniature Neutron Source Reactor. The MCNPX code was employed to model the 30 kW research reactor beam line. Six NBR sheets, each 2 cm thick, were placed at the exit of the beam line. These sheets were stacked sequentially to evaluate the shielding capabilities of the material against both neutrons and gamma rays emitted from the exit. The results obtained indicated that the flexible and cost-effective material could effectively shield against neutrons, while providing minimal protection against gamma rays (Gholamzadeh et al., 2024).

Researchers have recently conducted a study focusing on the significant effects of radioactive radiation on human health and the importance of radiation mapping. By using artificial neural networks trained with Monte Carlo data, they have developed a method to create high-resolution radiation maps. The researchers believe that this approach has the potential for practical implementation in real-world scenarios (Okabe et al., 2024). Based on the results of recent experimental and theoretical research, a series of simulation studies and calculations were conducted to determine the most suitable shield material. This involved the incorporation of various weight percentages of a new compound.

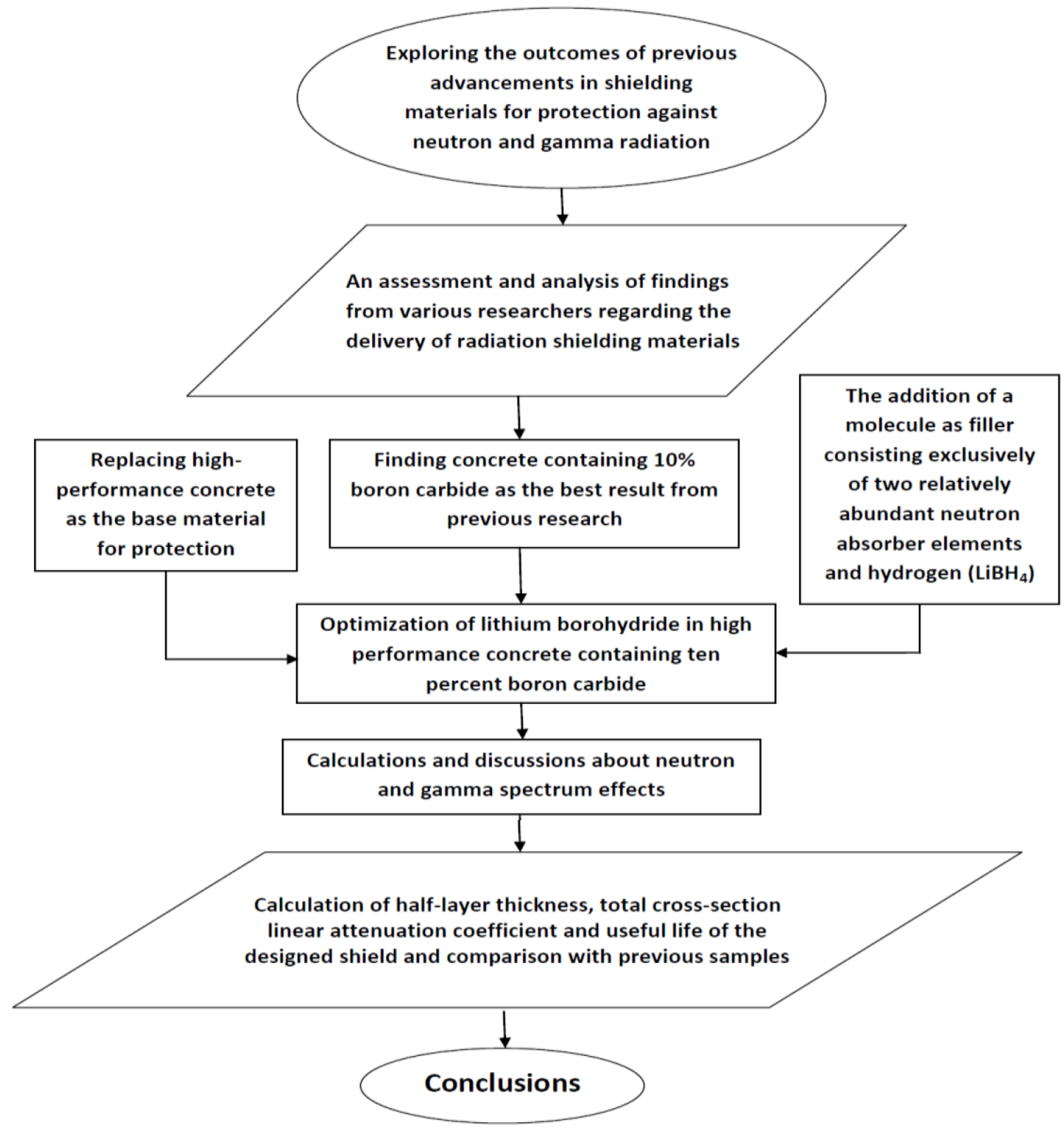

**Figure 1:** Flowchart of the research method.

## 2 Materials and Methods

The study initially focused on enhancing a novel single-layer material shield, measuring 5 cm in thickness, designed to provide protection against a broad range of neutron-gamma mixed radiation, including fast neutrons, without the inclusion of a moderator. Subsequent optimization efforts were directed towards refining this thickness.

Fast reactors are characterized by a relatively hard neutron spectrum, which results in a greater number of neutrons escaping. As a result, employing a fast reactor as a radiation source leads to increased interactions between neutrons and the shield. Additionally using of reactors as a source facilitates time-dependent analyses to assess the longevity of the shield. Therefore, utilizing a fast reactor as a source is deemed an advantageous option.

By evaluating and selecting the most suitable shield based on previous optimizations (concrete containing boron carbide), significant results can be achieved by replacing conventional concrete with HPC and gradually adding the new compound Lithium Borohydride ($LiBH_4$). The new compounds can be used in the future in industries that work with neutron beams, such as those related to nuclear fusion or accelerators. The flowchart of the research method is shown in Fig. 1.

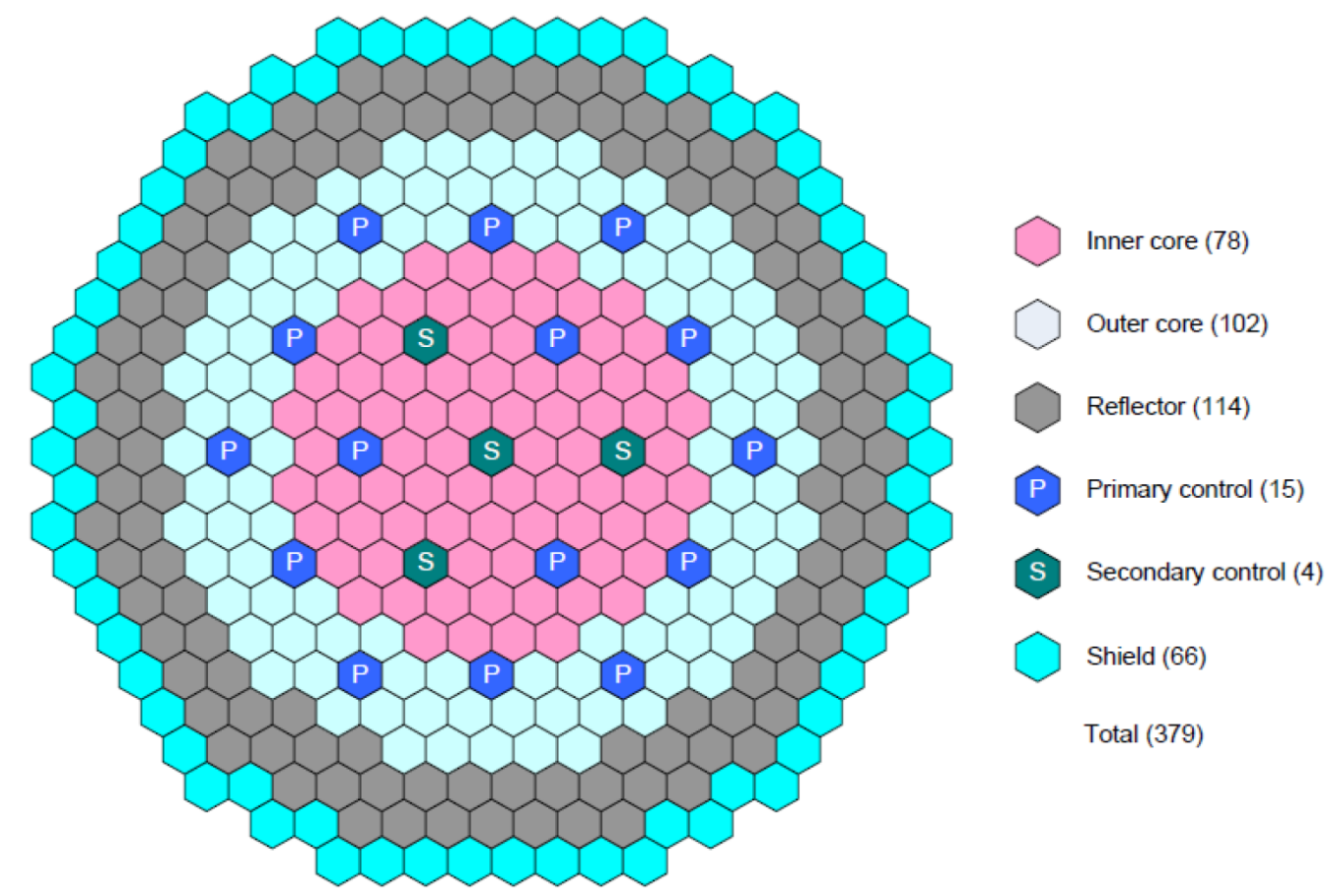

**Figure 2:** Radial layout of a MET-1000 core (Stauff et al., 2016).

Although the shield is being evaluated against a MET-1000 reactor, the objective is to design a radiation shield that can be applied in various scenarios. The source used is a fast reactor with a power of 1000 MWth containing a metal fuel known as MET-1000, as depicted in Fig. 2 (Stauff et al., 2016). The simulated reactor is a cylinder with a radius of 73.11 cm and a height of 85.82 cm. The overall density of the core was determined by averaging the volume ratios of its components. Additionally, there is a 10 cm thick layer of HT-9 surrounding the reactor core, which acts as a reflector (Stanković et al., 2010). The reactor operates at full power under critical conditions, producing both neutrons and gamma rays. The simulation is conducted using the MCNPX 2.7E code. The libraries utilized were not developed, but rather the ENDF70a to ENDF70k libraries. Neutrons were divided into 40 different energy groups. The calculations were conducted using 100,000 neutron histories over a total of 200 cycles, including 10 inactive cycles. This implies that a total of one hundred thousand neutron particles were considered in Monte Carlo simulations. Additionally, the analysis involved calculating changes in the multiplication factor over 200 steps, with the initial ten steps excluded from the calculations, resulting in a consideration of 190 steps. The shield is simulated as a rectangular cube with dimensions of 5 m in length, 5 m in wide and 5 cm in thickness. Also the shield volume is 5 m × 5 m × 5 cm =1.25 $m^3$. The distance between the shield and the source is 50 cm. A criterion for measuring flux changes after the shield is set up at 1 cm beyond the shield. Additionally, the cell volume in all flux measurements is 0.25 $m^3$ (5 m × 5 m × 1 cm = 0.25 $m^3$). Tables 1 and 2 present the ratios of the core and reflector compounds, respectively.

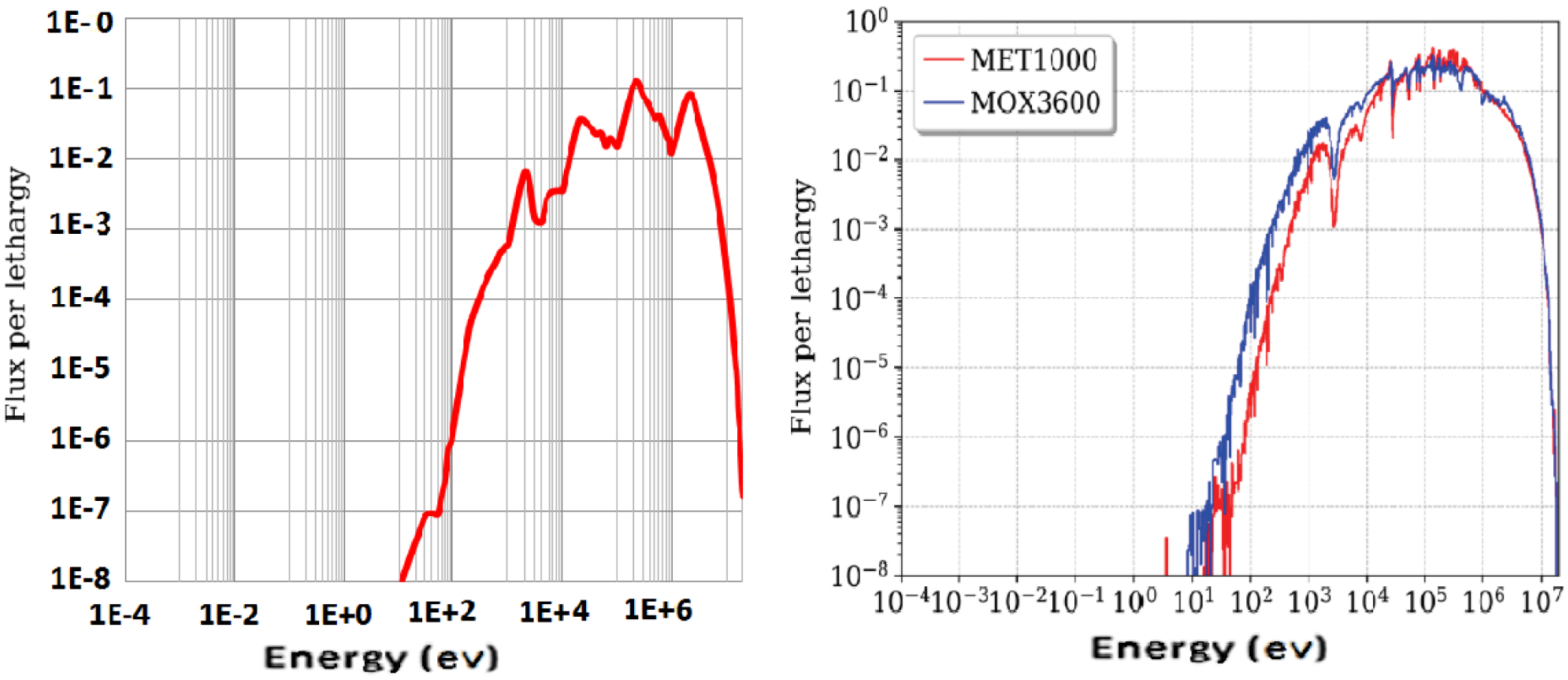

**Figure 3:** Comparison between the neutron spectra calculated in the simulated MET-1000 reactor (left side) and the main reference (Bostelmann et al., 2019) (right side).

In the first step, we evaluated five different materials based on the results of other optimization studies. These materials included an optimal alloy consisting of 5 wt% lead, 40 wt% iron, 50 wt% copper, and 5 wt% aluminum (Reda, 2016); concrete containing tungsten nanoparticles (Tekin et al., 2017); concrete containing 30wt% iron oxide (Gur et al., 2017); polyethylene containing 5 wt% boron carbide and 45 wt% tungsten oxide (Salimi et al., 2018); and concrete containing an optimal amount of 10 wt% boron carbide (Stanković et al., 2010). We also evaluated some conventional shields. After this evaluation, we selected concrete containing 10% boron carbide. In the second step, we considered the use of 10% boron carbide in HPC. When space is limited, HPC can reduce the thickness of the shield structure without compromising its shielding effectiveness. We evaluated these compounds to further enhance the shield.

1. HPC containing 10% boron carbide.

2. HPC containing 10% boron carbide and lithium borohydride ($LiBH_4$).

We evaluated of the shields under identical physical and geometrical conditions, considering the realistic densities of the materials. The examined materials are listed in Table 3. Additionally, Tables 4 and 5 display the weight ratios of concrete and polyethylene isotopes, respectively.

**Table 1:** Ratio of ingredients in the core.

| Material | Ratio | Material | Ratio |
|---|---|---|---|
| Na | 0.353 | Cr | 0.0327 |
| Pu-239 | 0.0351 | Mn | 0.00145 |
| U-238 | 0.0809 | Mo | 0.00155 |
| U-235 | 0.0057 | C | 0.001 |
| O | 0.247 | B-10 | 0.0012 |
| Fe | 0.236 | B-11 | 0.0031 |
| Ni | 0.0013 | - | - |

**Table 2:** Ratio in ingredients of the reflector.

| Material | Ratio | Material | Ratio |
|---|---|---|---|
| Si | 0.0259 | K | 0.00271 |
| Al | 0.00591 | Na | 0.0192 |
| Fe | 0.12625 | Sr | 0.00095 |
| Ca | 0.145 | O | 0.5085 |
| Mg | 0.00317 | H | 0.1555 |
| S | 0.00691 | - | - |

## 3 Calculations and Discussion

### 3.1 Benchmark results

In order to validate the simulated source, the neutron spectrum and $K_{eff}$ were compared with those of the MET-1000 reactor. The $K_{eff}$ of the simulated reactor was calculated to be $1.02 \pm 0.0007$, which is less than 1% different from the value given in the reference model (Stauff et al., 2016). The spectrum was found to be identical to that of the reference sample, and the average neutron flux in the entire core was determined to be $1.42 \times 10^{15}$ n.cm$^{-2}$.s$^{-1}$. Figure 3 shows a comparison between the neutron spectra calculated in the MET-1000 reactor and those in reference (Bostelmann et al., 2019).

**Table 3:** Compounds and elements studied in the first step.

| No. | Name | Material |
|---|---|---|
| 1 | Without Shield | Without shield |
| 2 | PE | Polyethylene |
| 3 | Conc. | Ordinary concrete |
| 4 | HPC | High performance concrete |
| 5 | Pb | Pure lead |
| 6 | Pb, Al, Cu, Fe | A compound including 5% Al, 40% Fe, 50% Cu and 5% Pb (Reda, 2016) |
| 7 | PE, $W_3O$, $B_4C$ | Polyethylene contains 5% $B_4C$, 45% $W_3O$ (Salimi et al., 2018) |
| 8 | Conc., $Fe_2O_3$ | Concrete contains 30% $Fe_2O_3$ (Gur et al., 2017) |
| 9 | Conc., $W_3O$ | Concrete contains tungsten nanoparticles (Tekin et al., 2017) |
| 10 | Conc., $B_4C$ | Concrete contains 10% $B_4C$ (Stanković et al., 2010) |
| 11 | HPC, $B_4C$ | High performance concrete contains 10% $B_4C$ |
| 12 | HPC, $B_4C$, $Li_2O$ | High performance concrete contains 10% $B_4C$, $Li_2O$ |
| 13 | HPC, $B_4C$, $LiBH_4$ | High performance concrete contains 10% $B_4C$, $LiBH_4$ |

**Table 4:** Ratio of concrete isotopes (Busch et al., 2009).

| No. | Element | Element weight fraction in concrete | Isotope natural abundance | Isotope weight percent in concrete |
|---|---|---|---|---|
| 1 | Si | 0.337 | Si-28 (92.27 %) | Si-28 (31.1 %) |
| | | | Si-29 (4.68 %) | Si-29 (01.58 %) |
| | | | Si-30 (3.05 %) | Si-30 (01.03 %) |
| 2 | Al | 0.337 | Si-27 (4.68 %) | Si-27 (3.4 %) |
| 3 | Fe | 0.014 | Fe-54 (5.84 %) | Fe-54 (0.082 %) |
| | | | Fe-56 (91.68 %) | Fe-56 (1.28 %) |
| | | | Fe-57 (2.17 %) | Fe-57 (0.028 %) |
| | | | Fe-58 (0.31 %) | Fe-58 (0.00434 %) |
| 4 | Ca | 0.044 | Ca-40 (100 %) | Ca-40 (4.4 %) |
| 5 | Na | 0.029 | Na-23 (100 %) | Na-23 (2.9 %) |
| 6 | O | 0.532 | O-16 (99.759 %) | O-16 (53.07 %) |
| | | | O-17 (0.037 %) | O-17 (1.97 %) |
| | | | O-18 (0.204 %) | O-18 (10.85 %) |
| 7 | H | 0.01 | H-1 (≈100 %) | H-1 (1 %) |
| - | Total | 1.00 | - | - |

$$S = \sqrt{\frac{1}{N-1}\sum_{i=1}^{N}(K_i - \bar{K})^2} \tag{1}$$

Because the computation of $K_{eff}$ and the chain reactions depend on the rest of the parameters, the standard deviation also applies to them.

### 3.2 *Evaluation of shields against neutron radiation*

The average flux after passing through the 5 cm thick shield is shown in Fig. 4. The graphs are organized based on the materials' effectiveness in reducing flux. The lowest neutron flux was observed when using HPC containing boron carbide. The neutron spectrum after passing through the three different materials is shown in Fig. 5. As can be seen from the graph, the addition of boron carbide is more effective for low-energy neutrons (below 10 keV).

### 3.3 *Evaluation of Shields against Gamma Radiation*

The results for all types of shields against gamma radiation are shown in Fig. 6. The gamma spectrum after three layers of shielding is shown in Fig. 7.

Although the total gamma flux after the shields was not equal, the intended shields effectively attenuated the gamma rays and exhibited similar fluctuations in relation to the spectrum. Different shields do not equally reflect neutrons; therefore, the ratio of neutron flux after and before the shields is not a suitable criterion for comparison. The main criterion for a shield is the flux rate after the shield. Tungsten-containing shields have a relatively positive impact on the effectiveness of gamma shields. However, the reactions $^{182}$W(n,2n)$^{181}$W, $^{186}$W(n,2n)$^{185}$W, and $^{186}$W(n,3n)$^{184}$W (Konobeyev et al., 2019) showed poor performance against fast neutrons.

**Table 5:** Ratio of polyethylene isotopes (Busch et al., 2009).

| No. | Element | Element weight fraction in polyethylene | Isotope natural abundance | Isotope weight percent in polyethylene |
|---|---|---|---|---|
| 1 | Hydrogen | 0.143 | H-1 (≈100 %) | H-1 (≈14.3 %) |
| 2 | Carbon | 0.857 | C-12 (98.89 %)<br>C-13 (1.11 %) | C-12 (84.75 %)<br>C-13 (0.95 %) |
| - | Total | 1.00 | - | 1.00 |

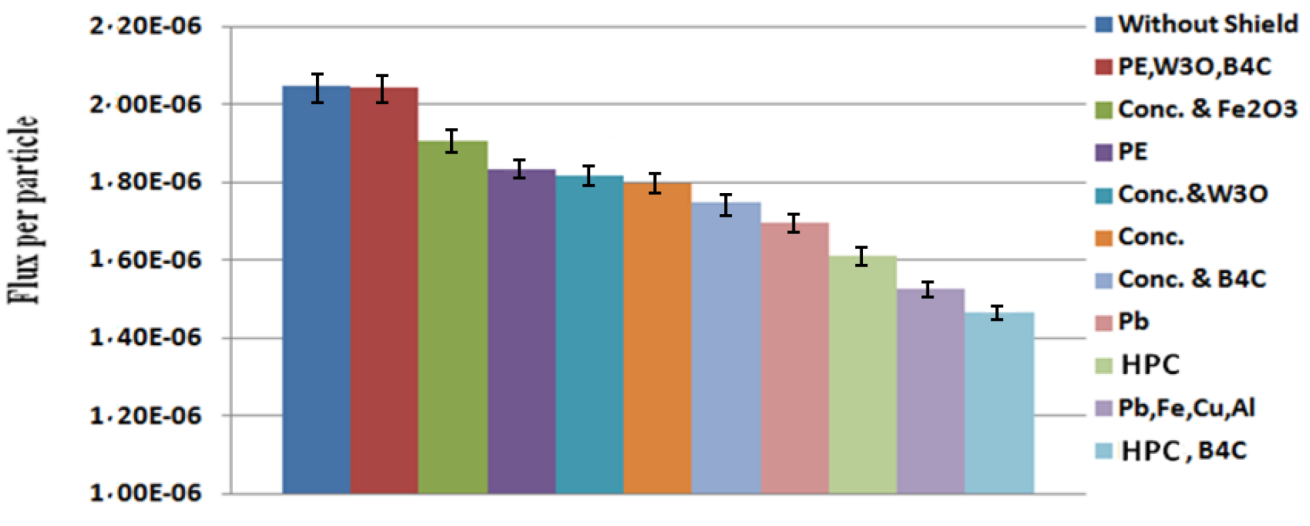

**Figure 4:** Total neutron flux after different types of shields with a thickness of 5 cm.

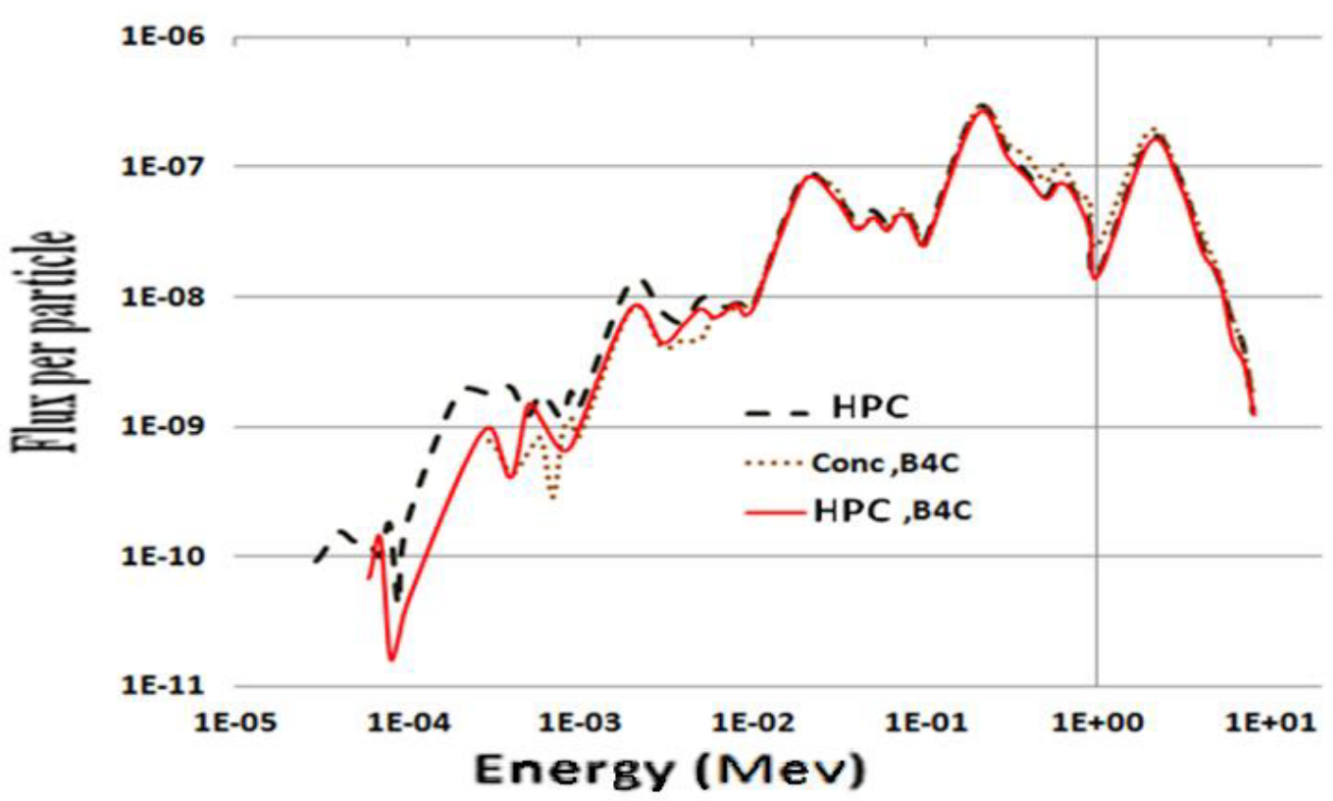

**Figure 5:** Neutron spectrum after three different materials with a thickness of 5 cm.

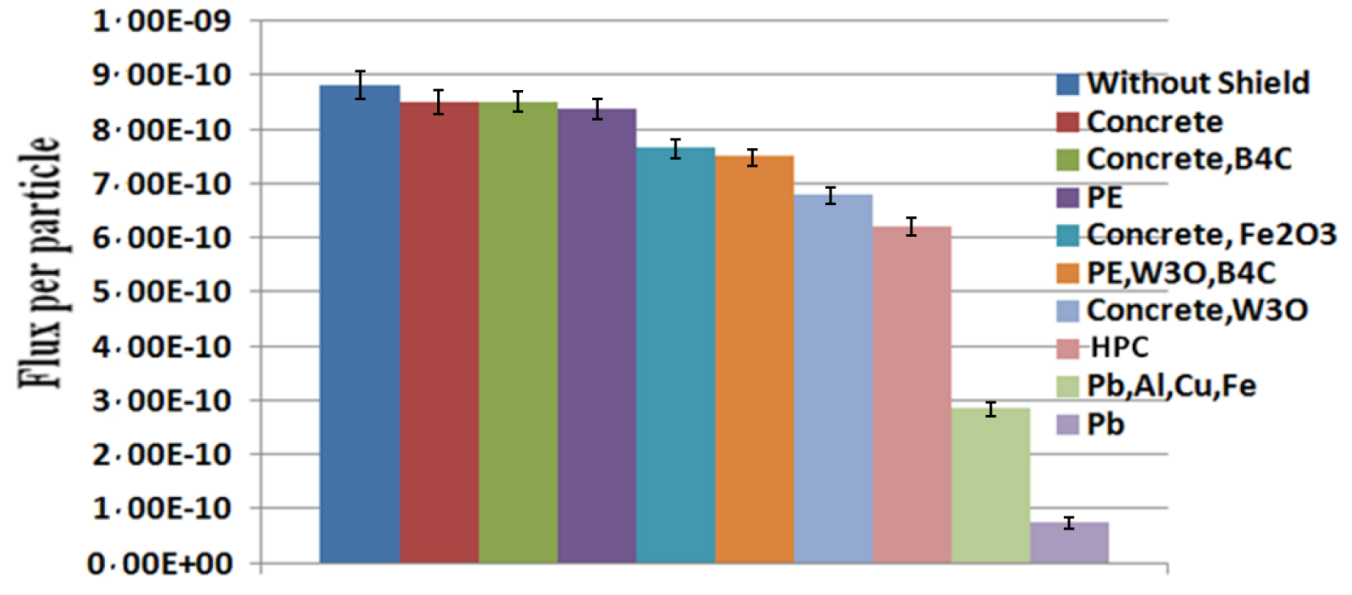

**Figure 6:** Total gamma flux after using different types of shields with a thickness of 5 cm.

### 3.4 *Behavior of High Performance Concrete and Lithium Borohydride (LiBH$_4$)*

High-performance concrete: This type of concrete has superior strength and long-term performance. The typical composition of a proprietary HPC mix can be seen in Table 6 (Laiblová et al., 2019).

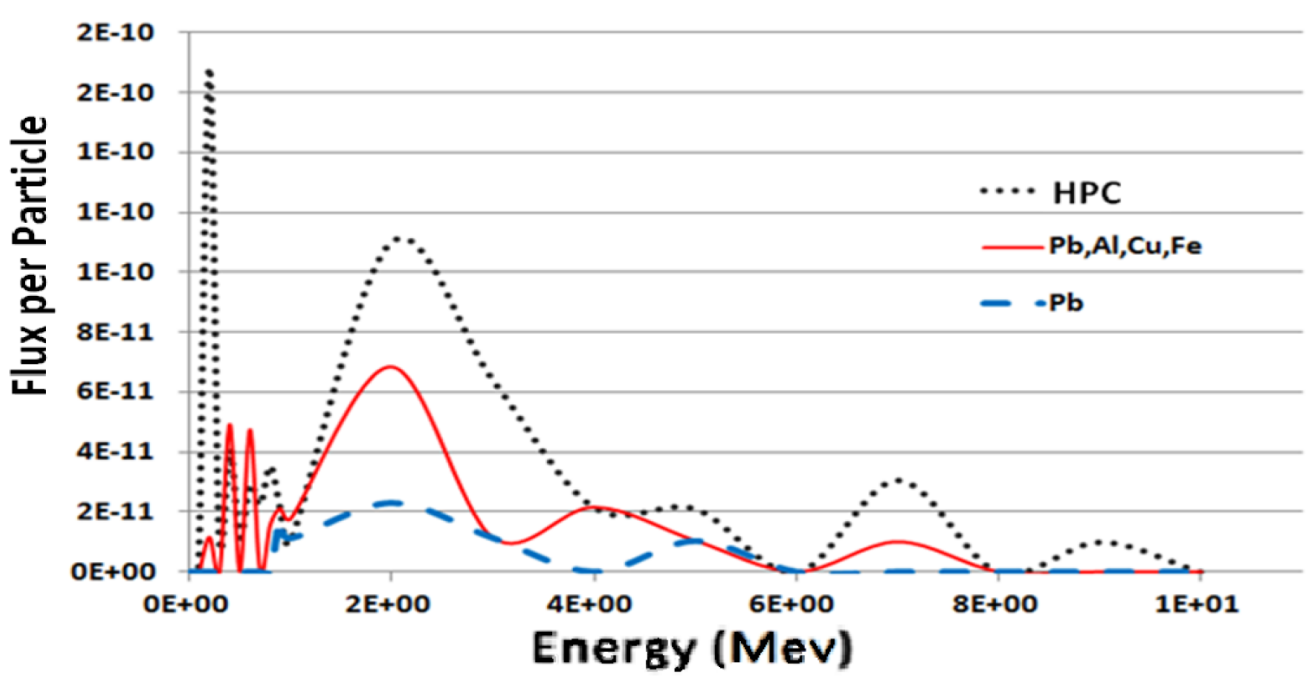

**Figure 7:** Gamma spectrum after passing through three different materials with a thickness of 5 cm.

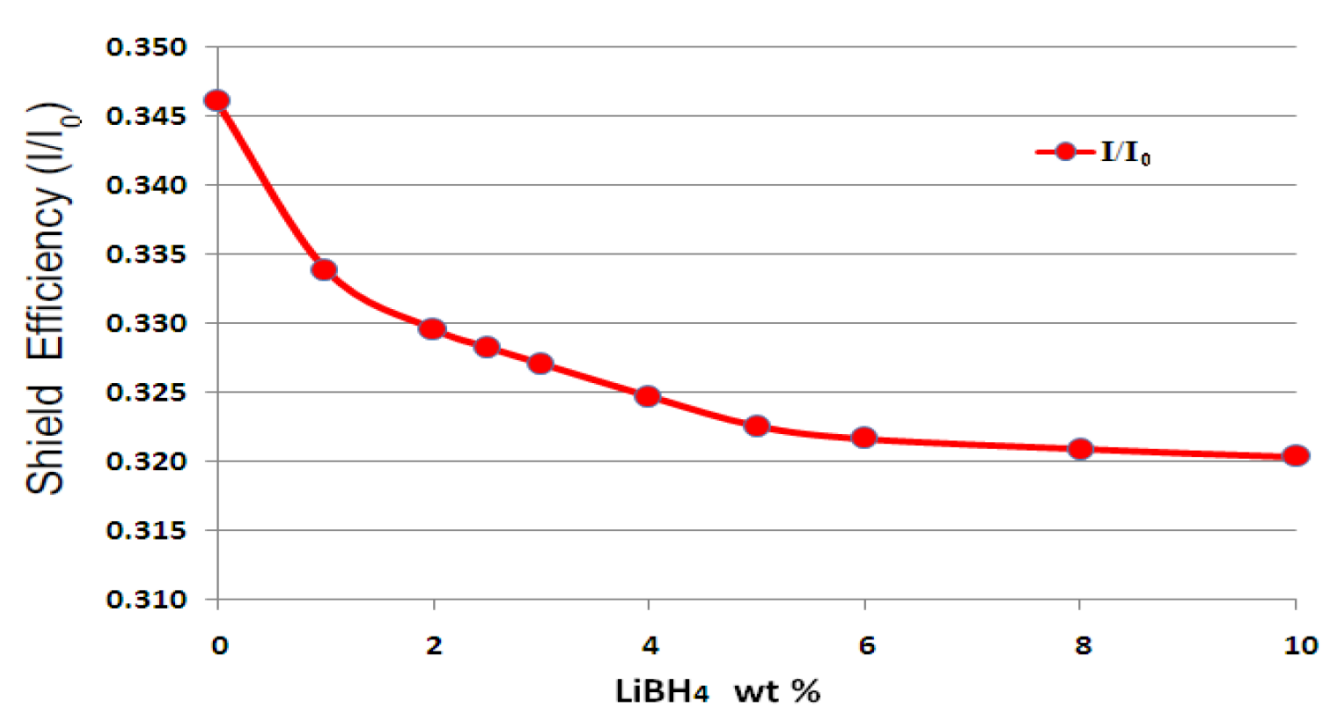

**Figure 8:** Variation of neutron flux relative to LiBH$_4$ content in the HPC shield containing 10% B$_4$C and 5cm thickness: The neutron flux decreases again after the shield by increasing the LiBH$_4$ content.

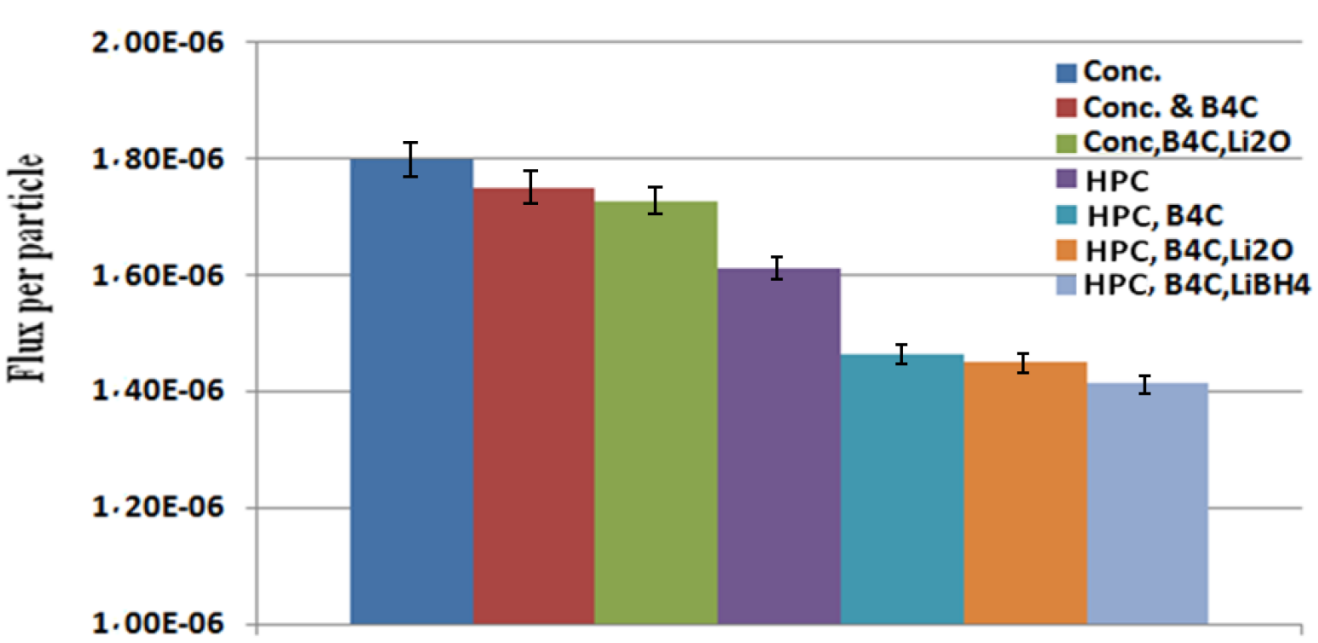

**Figure 9:** Comparison of the calculated neutron flux after passing through new and conventional shields.

New compounds can be developed to enhance shield effectiveness, so we decided to focus on lithium compounds as well. From the array of lithium-containing chemicals available, we chose lithium borohydride (LiBH$_4$) as our primary compound. $LiBH_4$ was selected because each molecule contains one boron atom (an absorber) and four hydrogen atoms (moderators), making it a promising candidate. Additionally, lithium borohydride is readily accessible. Our research showed that incorporating lithium borohydride in shield compounds led to a significant exponential increase in effectiveness, even with optimized amounts of boron carbide. To determine the ideal proportion of $LiBH_4$, we conducted experiments with different ratios of $LiBH_4$ and boron carbide in the shield compound. To determine the optimal content of lithium borohydride, we tested various percentages of $LiBH_4$ with shields of varying thicknesses. Figure 8 illustrates that as the percentage of lithium borohydride in the shield increased, the neutron flux decreased.

**Table 6:** Composition of the HPC mix (Laiblová et al., 2019).

| HPC | | OC | |
|---|---|---|---|
| Component | kg.m$^{-3}$ | Component | kg.m$^{-3}$ |
| Technical sand | 979 | Sands | 1150 |
| Cement I 42.5R | 693 | Cement CEM II/B-M (S-LL) 23.5 | 360 |
| Quartz floor | 332 | Gravel | 810 |
| Silica fume | 178 | - | - |
| Superplasticizer | 29.6 | Superplasticizer | 2.7 |
| Water | 174 | Water | 155 |
| Total | 2385.6 | Total | 2477.7 |

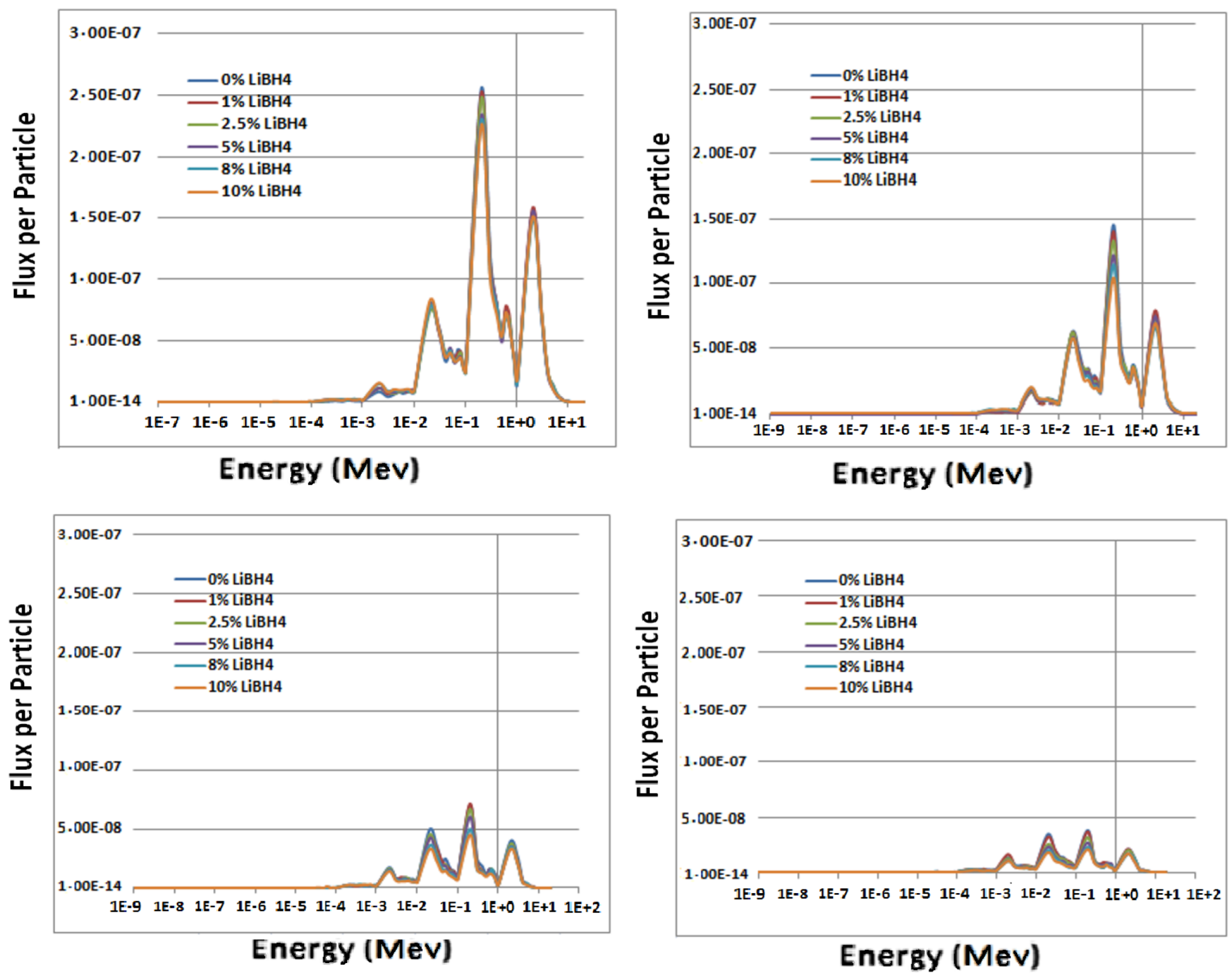

**Figure 10:** Spectra of neutrons passing through shields with varying thicknesses (5, 10, 15, and 20 cm) and different weight percentages of $LiBH_4$.

The bar chart in Fig. 9 shows that the total neutron flux is lower after passing through the new shields compared to the previous shields. The spectra of the six shields at different thicknesses (5, 10, 15, and 20 cm) are shown in Fig. 10.

The transfer of neutron radiation can be classified into two types of shields based on the neutron spectrum. Figure 11 shows the results.

As shown in the graph, the addition of $LiBH_4$ is more effective in preventing neutrons in the energy range of 100 keV to 1 MeV. Additionally, the transmission ratio of gamma radiation through three main types of shields was classified based on the gamma spectrum in two regions. The amount and range of gamma radiation passing through the shield are determined not only by density but also by the effectiveness of fillers such as $B_4C$ and $LiBH_4$. The role of density is significant, with HPC responding better to gamma radiation than ordinary concrete. The impact of fillers containing elements like boron and lithium on electromagnetic waves has already been studied (Yorgun et al., 2019).

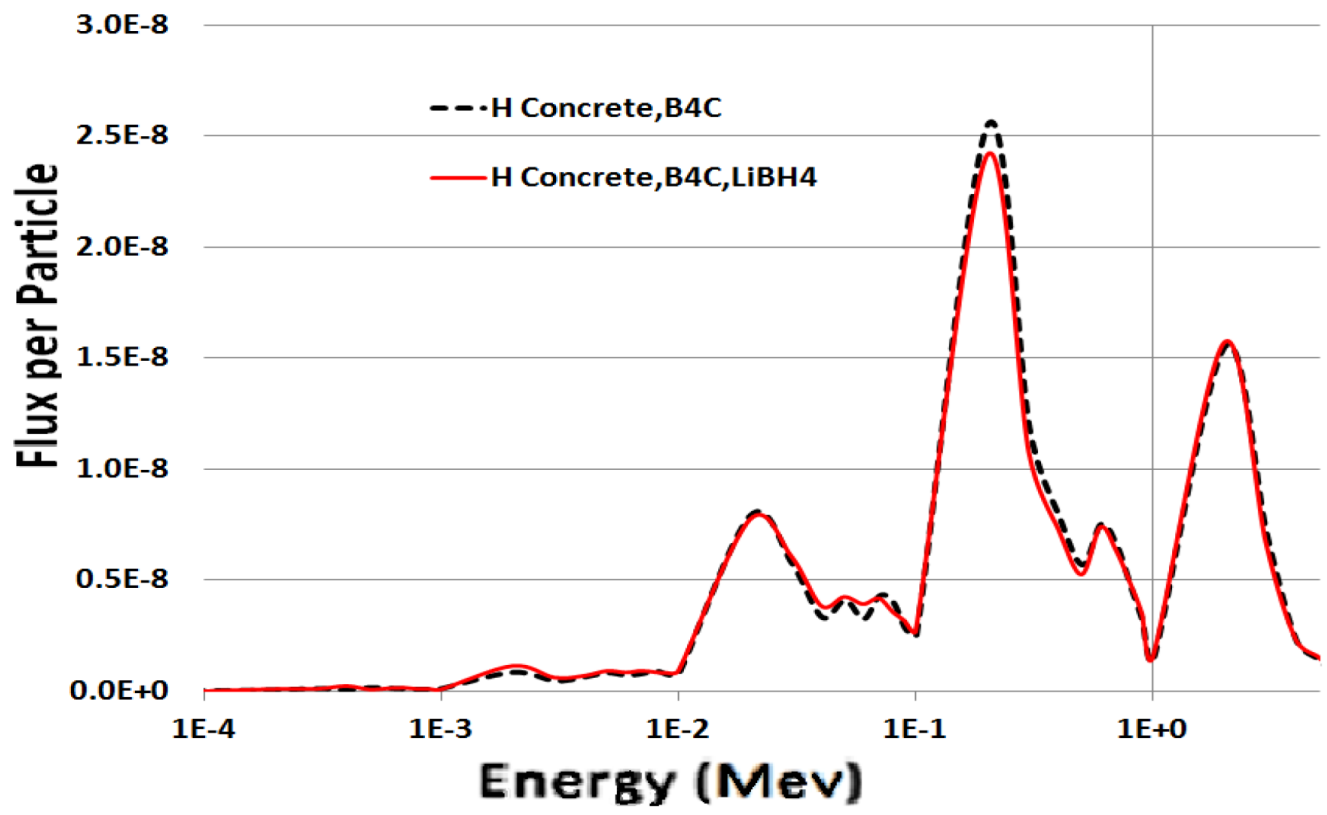

**Figure 11:** Two magnifications of the neutron spectrum after two types of shields with a thickness of 5 cm.

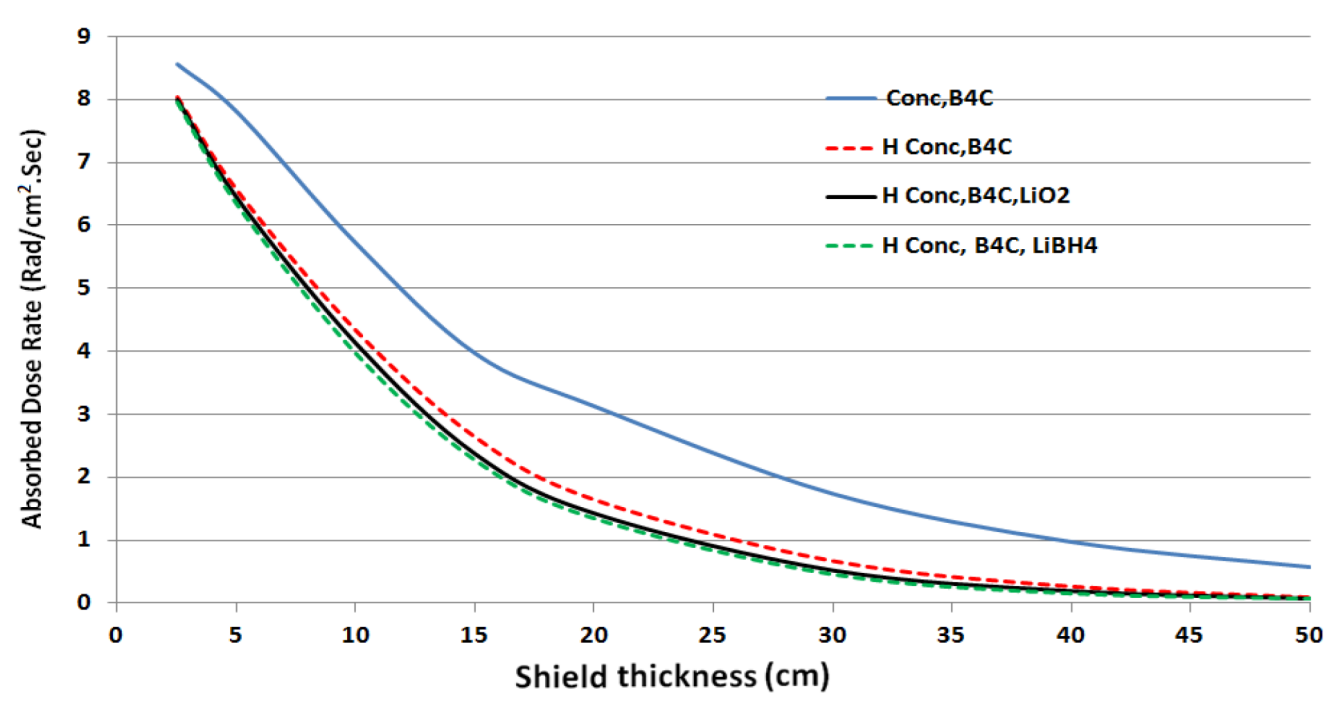

**Figure 12:** Changes in absorbed dose rate relative to variations in shield thickness.

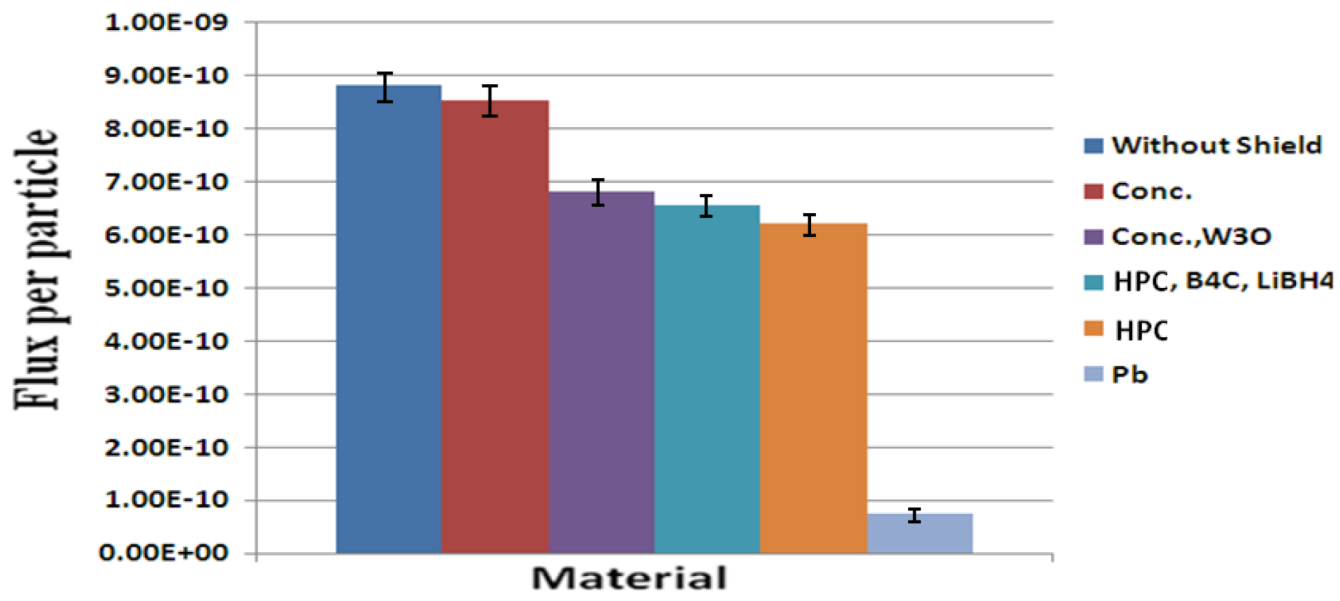

**Figure 13:** Comparison of gamma flux after the new shield and some previous shields.

### 3.5 Calculation of other parameters

Calculation of other parameters involves determining the half-value layer (HVL) by gradually increasing the shield thickness using the MCNPX code. The linear attenuation coefficient ($\lambda_l$) is found using the equation HVL=0.693 $\lambda_l$. It is assumed that the total cross-section ($\Sigma_t$) is inversely related to the diffusion length. Tables 7 and 8 show the HVL, total cross-section, and linear attenuation coefficient of the shield designed to protect against gamma and neutrons.

Table 9 presents the neutron and gamma flux results obtained from evaluating of the key shielding materials used in this study.

**Table 7:** Calculated data for several shields against neutron radiation.

| Material | $\Sigma_t$ ($cm^{-1}$) | Half value layer (cm) | $\lambda_l$ (cm) |
|---|---|---|---|
| HPC, $B_4C$ , $LiBH_4$ | 2.09E-01 | 3.32 | 4.79E+00 |
| Conc.,$B_4C$ | 1.55E-01 | 4.471 | 6.46E+00 |

**Table 8:** Calculated data for several shields against gamma radiation.

| Material | $\Sigma_t$ ($cm^{-1}$) | Half value layer (cm) | $\lambda_l$ (cm) |
|---|---|---|---|
| HPC, $B_4C$ , $LiBH_4$ | 2.81E-01 | 2.47 | 3.56E+00 |
| Conc.,$B_4C$ | 1.31E+00 | 0.53 | 7.66E-01 |

As expected, the designed shield performed better against neutrons than the concrete containing boron carbide. However, the results indicate that the lead shield is more effective against gamma radiation. It is important to determine if the optimal thickness of the shield against neutron radiation also provides satisfactory results against gamma radiation. Therefore, the thickness of the shield against neutrons should be optimized first.

### 3.6 Shield thickness optimization

The difference in neutron flux and absorbed dose becomes more evident as shield thickness gradually increases, highlighting the distinct impacts of different shield types. Figure 10 illustrates that the characteristics of the particles and their corresponding spectrum exhibit minimal variation as they traverse the shield; consequently, the absorption dose was deemed to be directly proportional to the alterations in flux. Figure 12 depicts the absorbed dose rate after passing through four materials of varying thicknesses.

The effect of material change increases with thickness. For example, to allow 10% of the neutron flux to pass through, approximately 25 cm of HPC containing $B_4C$ (boron carbide) and $LiBH_4$ (lithium borohydride) would be needed. In contrast, 40 cm of regular concrete with boron carbide would be required, representing a 40% reduction in thickness. Additionally, increasing the thickness of HPC with 10% $B_4C$ and 5% $LiBH_4$ by more than 30 cm has minimal effect. Therefore, the optimal shield thickness is determined to be 30 cm.

### 3.7 Effect of Monolayer Shielding Against Gamma Rays

In this phase, we calculated the gamma flux value after optimizing the shield and compared it to previous measurements, depicted in Fig. 13. The figure indicates that while the total gamma flux post-implementation of the new shield may not be lower than all previous shields (especially when compared to lead), its performance is deemed satisfactory. Figure 14 demonstrates the changes in gamma flux with an increase in shield thickness.

**Table 9:** Calculated neutron and gamma fluxes per particle after 5 cm thickness of various shields. The values are reported per particle.

| Material | Neutron flux after 5 cm shield | Gamma flux after 5 cm shield |
|---|---|---|
| Conc. and $W_3O$ (Tekin et al., 2017) | 1.82E-06 | 6.81E-10 |
| PE, $W_3O$, $B_4C$ (Salimi et al., 2018) | 2.05E-06 | 7. 5E-10 |
| Pb, Fe, Cu, Al (Reda, 2016) | 1.53E-06 | 2.85E-10 |
| Conc. and $B_4C$ (Stanković et al., 2010) | 1.75E-06 | 8.5E-10 |
| Conc. and $Fe_2O_3$ (Gur et al., 2017) | 1.91E-06 | 7.67E-10 |
| Pb | 1.70E-06 | 7.39E-11 |
| HPC | 1.61E-06 | 6.20E-10 |
| HPC, $B_4C$ | 1.47E-06 | 6.04E-10 |
| HPC, $B_4C$, $Li_2O$ | 1.43E-06 | 6.01E-10 |
| HPC, $B_4C$, $LiBH_4$ | 1.41E-06 | 5.97E-10 |

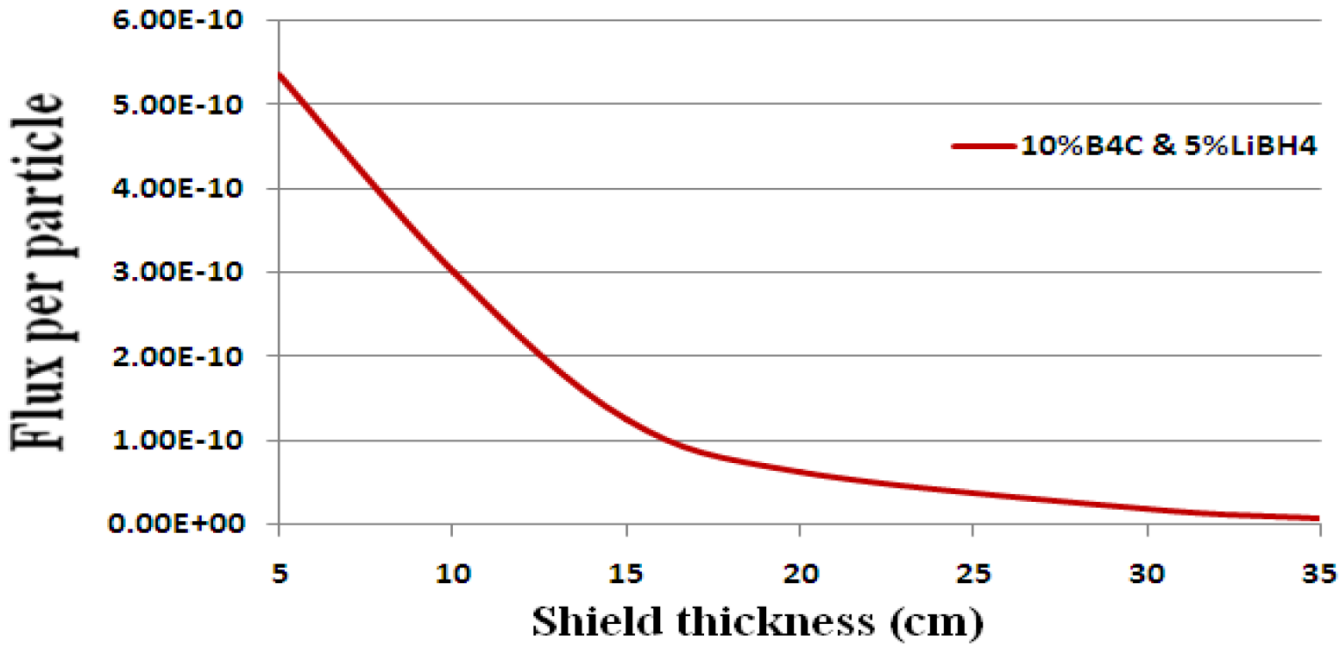

**Figure 14:** Changes in gamma flux after shielding relative to a gradual increase in shield thickness.

Now, let's consider the optimal thickness of the shield, which is made of concrete containing boron carbide and $LiBH_4$, against neutron radiation. It has been observed that a shield with a thickness of 30 cm can prevent more than 92% of the total gamma radiation. The calculation was also repeated for a shield with a thickness of 35 cm, and it was found that a 35 cm thick shield completely blocks the passage of gamma radiation. Calculations show that if this type of shield is used with a thickness of 30 cm or more, there is no need to use lead as a second layer for gamma shielding. Additional calculations were conducted for the shield using various thicknesses, and it was determined that altering the weight percentage of lithium borohydride had no impact on the gamma attenuation coefficient.

Recent findings suggest that, in addition to increasing density, various intriguing factors can also play a role in enhancing the efficacy of a gamma shield. It is currently understood that melanin in the skin has the ability to provide protection against radiation through its interaction with electromagnetic radiation (Xie et al., 2024). This protective mechanism can also be facilitated by distinct chemical reactions that have been demonstrated through artificial modeling.

### *3.8 Shield quality over time*

The interaction between the neutrons and shield materials can have a significant impact, even at low levels. Therefore, it is necessary to calculate the consumption of the main neutron absorber isotopes when optimizing new shields under various conditions. However, if a shield is positioned close to a powerful source, such as the Met-1000 reactor, the flux inside the shield cannot be ignored. To assess the production of hazardous isotopes resulting from the interaction between the shield and neutrons, time-dependent calculations are helpful.

The importance of radiation protection and environmental health cannot be overstated. As the transition to different reactor fuel types is pursued for economic reasons, experts are also anticipating shifts in nuclear waste management in response to pressing concerns for environmental and public health (Lotfalian et al., 2024).

To evaluate the lifespan of a concrete shield containing 10% boron carbide and 5% $LiBH_4$ with a thickness of 30 cm, the quality of the shield was assessed for 180 days after the reactor reached full power. Time-dependent calculations were performed using the BURN card to determine the reduction amount of B-10 and Li-6, which are the main absorbers. The results are presented in Table 10.

Although the neutron flux inside the shield is low, some absorbing isotopes, such as B-10 and Li-6, have reacted with neutrons and decreased. The consumption of B-10 and Li-6 after 180 days, compared to their initial values, was 0.32% and 0.054%, respectively.

**Table 10:** Changes (consumption) in adsorbent isotopes in the shield over time.

| Time (days) | B-10 (g) | Li-6 (g) |
|---|---|---|
| 0 | 1.242E+05 | 3.711E+03 |
| 80 | 1.240E+05 | 3.710E+03 |
| 180 | 1.238E+05 | 3.709E+03 |

## 4 Conclusions

In this study, we examined various shields against a wide spectrum of neutron-gamma mixed fields. We assessed different conventional shields and considered the optimized results from previous studies as potential shielding materials. The primary findings indicated that concrete with 10%wt of boron carbide offered the best performance against neutron radiation. By substituting High-Performance Concrete (HPC) with regular concrete and gradually adding lithium borohydride, we observed an en-hancement in shield quality, even with the optimal 10 wt% concentration of boron carbide in the shield compound.

The ideal weight percentage of lithium borohydride in the designed shield was determined to be 5 wt%. Under these conditions, the new shielding material (HPC containing 10% $B_4C$ and 5% $LiBH_4$) could reduce volume by 40.0% compared to concrete with boron carbide. Boron carbide is more effective for low-energy neutrons (below 10 keV), while $LiBH_4$ is more effective for neutrons in the energy range of 100 keV to 1 MeV. As neutrons travel through the shield, many fast neutrons are converted into thermal neutrons, decreasing their energy and increasing the absorption cross-section. In such cases, a moderator is often unnecessary. Due to the high percentage of fast neutrons in the source, as it is a fast reactor, lithium borohydride was found to be more effective than boron carbide as a shield against fast neutrons. Consequently, when these neutrons pass through a shield containing light elements, they become thermal neutrons, increasing the microscopic capture cross-section of absorption across the entire shield composition. However, incorporating fillers like $B_4C$ and $LiBH_4$ can significantly improve shielding effectiveness. While HPC containing $B_4C$ and $LiBH_4$ was effective against neutrons, a shield thickness of 30 cm blocked 95% of the total neutron spectrum and 92% of the total gamma spectrum. Therefore, gradually increasing the shield's thickness to 30 cm or more eliminates the need for a second layer of the shield to protect against gamma radiation.

Despite the substantial improvement in shield quality achieved by including compounds with neutron-absorbing isotopes like B-10 and Li-6, the depletion of these isotopes was minimal. After 180 days of full power reactor operation, only 0.32% of B-10 and 0.054% of Li-6 were consumed. This suggests that the shield's quality remained relatively constant over time, indicating a long expected useful life.

Lithium borohydride demonstrates promising properties for neutron radiation shielding and shows potential for novel applications and optimization. Moreover, $LiBH_4$ offers opportunities for customization and optimization, allowing for tailored applications to ensure optimal radiation shielding performance.

## Conflict of Interest

The authors declare no potential conflict of interest regarding the publication of this work.

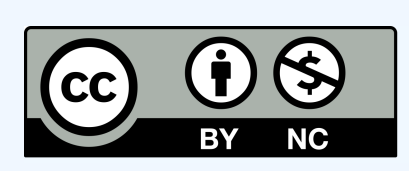

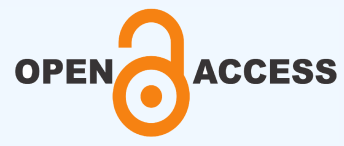